\documentclass[a4paper,11pt]{article}
\usepackage{pos}

\title{Studying QGP transport properties in a concurrent minijet+hydro framework}

\author*[a]{Charles Gale}
\author[a]{Sangyong Jeon}
\author[b,c,d]{Daniel Pablos}
\author[e]{Mayank Singh}

\affiliation[a]{Department of Physics, McGill University,\\
  3600 University Street, Montreal, QC, H3A 2T8, Canada }

\affiliation[b]{Departamento de Fisica, Universidad de Oviedo\\
Avda. Federico Garcia Lorca 18, 33007 Oviedo, Spain}

\affiliation[c]{Instituto Universitario de Ciencias y Tecnologias Espaciales de Asturias (ICTEA)\\
Calle de la Independencia 13, 33004 Oviedo, Spain}

\affiliation[d]{INFN, Sezione di Torino, via Pietro Giuria 1, I-10125 Torino, Italy}

\affiliation[e]{School of Physics \& Astronomy, University of Minnesota\\ Minneapolis, MN 55455, USA}

\emailAdd{gale@physics.mcgill.ca}
\emailAdd{jeon@physics.mcgill.ca}
\emailAdd{daniel.pablos.alfonso@to.infn.it}
\emailAdd{singh547@umn.edu}

\abstract{Minijets are ubiquitous in heavy-ion collision experiments. However, they are often excluded from the hydrodynamic simulations of QGP as they do not thermalize at short time scales and are not treated as part of the collective medium. Using a concurrent jet+hydro framework, we show that the minijets could account for a significant portion of particle multiplicity. Therefore, the energy deposition from minijet-medium interactions can substantially modify the QGP transport properties inferred from model-to-data comparisons.}

\FullConference{HardProbes2023\\
 26-31 March 2023\\
 Aschaffenburg, Germany\\}

\begin{document}
\maketitle

\section{Introduction}
There is a wide consensus that a deconfined state of nuclear matter, the quark-gluon plasma (QGP), is created in ultrarelativistic heavy-ion collisions~\cite{Baym:2016wox}. Relativistic dissipative hydrodynamics has been very successful in describing the low-momentum collective motion of the QGP~\cite{Gale:2013da}. Heavy-ion collisions produce numerous moderate-energy jets (minijets) by hard scatterings at early times~\cite{Arnold:2002zm}. These minijets do not thermalize and cannot be effectively treated as part of the fluid. They traverse the medium and deposit energy and momentum through interactions with the QGP. Minijets can account for a significant portion of total multiplicities and act as significant sources of local energy-momentum fluctuations~\cite{Pablos:2022piv}.

Minijet evolution can be affected by other minijets whose wake lies in its path. They can give rise to Mach wakes in the fluid and significantly change the evolutionary history of the medium. Consequently, a consistent treatment of minijet dynamics requires a concurrent jet+hydro framework where the minijets and the bulk medium inform each other's evolution.

Here, we study the effect of minijets on the extraction of QGP transport properties. We describe our framework in section~\ref{sec:framework}, look at the modification to hydro evolution in section~\ref{sec:hydro} and discuss our findings in section~\ref{sec:results}.

\section{Our framework}\label{sec:framework}
The bulk QGP is initialized using the IP-Glasma model~\cite{Schenke:2012wb}. 
This describes physics below the saturation scale $Q_s$. The color fields in IP-Glasma are evolved for 0.4 fm/c after the collision. The minjets are initialized by employing the hard processes in the PYTHIA8 framework~\cite{Sjostrand:2007gs}. The space-time positions of the binary nucleon-nucleon collisions used for IP-Glasma and PYTHIA are the same.

The bulk medium is evolved using $3+1$ D viscous hydrodynamic solver MUSIC~\cite{Schenke:2010nt}. The minijet energy loss is fed into the bulk evolution via a source term~\cite{Pablos:2022piv}. The energy-momentum tensor $T^{\mu\nu}_{\text{hydro}}$ evolves as
\begin{equation}
    \partial_\mu T^{\mu\nu}_{\text{hydro}} = J^\nu,
\end{equation}
where the source term $J^\nu$ is the momentum deposition convoluted with a Gaussian with width $\sigma_x$ in $x$ and $y$ directions, and width $\sigma_\eta$ in the rapidity direction,
\begin{equation}
    J^\nu = \sum_i \frac{\Delta P_i^\nu}{\Delta\tau(2\pi)^{3/2}\sigma_x^2\sigma_\eta\tau}e^{-\frac{\Delta x_i^2+\Delta y_i^2}{2\sigma_x^2}}e^{-\frac{\Delta\eta_i^2}{2\sigma_\eta^2}}.
\end{equation}
Here, $\tau$ is proper time, $\Delta\tau$ is the size of evolution time-step, and $(\Delta x_i, \Delta y_i, \Delta \eta_i)$ is the spatial distance to the $i^{\text{th}}$ momentum deposition $\Delta P^\nu_i$.

The energetic partons travel a finite distance in the QGP before stopping. The minijet energy loss is treated within the hybrid strong-weak coupling model~\cite{Casalderrey-Solana:2014bpa}. The parton splittings are governed by the weakly coupled perturbative interactions while the minijet-QGP interaction is governed by the strongly coupled energy loss formula~\cite{Chesler:2014jva}. The energy lost per unit length is given as
\begin{equation}
    \left.\frac{dE}{dx}\right|_{\text{strongly coupled}} = -\frac{4}{\pi}E_{\text{in}}\frac{x^2}{x_{\text{stop}}^2}\frac{1}{\sqrt{x^2_{\text{stop}}-x^2}} \, .
\end{equation}
Here, $E_{\text{in}}$ is the initial parton energy and $x_{\text{stop}}$ is the stopping distance. In the strongly coupled limit, the stopping distance can be obtained from holographic calculations~\cite{Gubser:2008as,Hatta:2008tx} as
\begin{equation}
    x_{\text{stop}}^{\text{AdS/CFT}} = \frac{1}{\kappa_i T}\left(\frac{E}{T}\right)^{1/3},
\end{equation}
where $\kappa_i$ is a species dependent parameter and $T$ is temperature.

The bulk medium is hadronized using the Cooper-Frye formalism~\cite{Cooper:1974mv} and the surviving minijets are hadronized using the Lund string model in PYTHIA. The partons that never crossed the freezeout hypersurface and were never quenched are hadronized using a corona color neutralisation model (CCN) where the original parton color is preserved, as in vacuum. The quenched partons have their colors randomized and are hadronized along with the medium in the local thermal color neutralisation model (LTCN). All the hadrons undergo cascading via UrQMD~\cite{Bass:1998ca}.

\section{Modification to hydro}\label{sec:hydro}
\begin{figure*}
    \centering
    \includegraphics[width=0.45\textwidth]{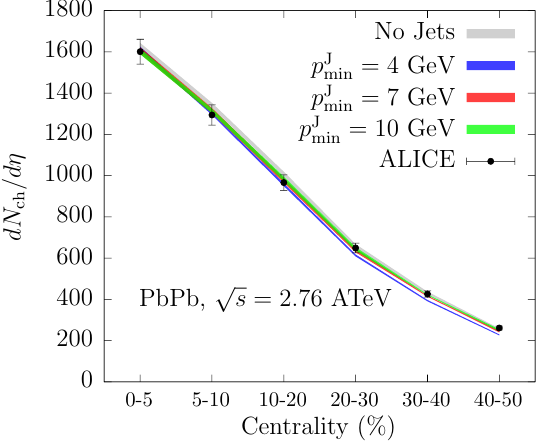}
    \includegraphics[width=0.42\textwidth]{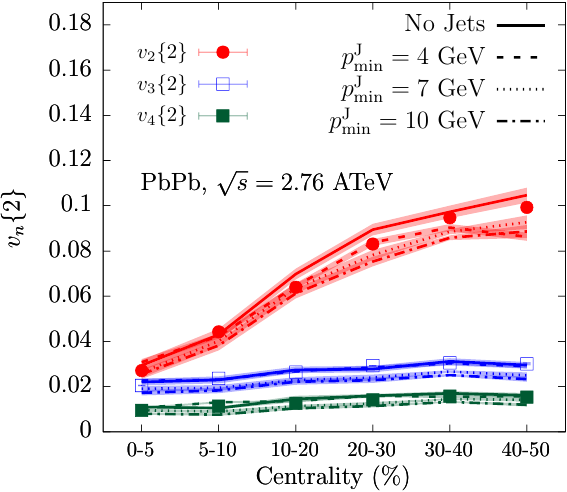}
    \caption{The charged hadron spectra (left) and $v_2$ (right) as a function of centrality for different choices of $p^{\text{J}}_{\text{min}}$, with suitably adjusted $s_{\text{factor}}$ and $\eta/s$.}
    \label{fig:spectra_v2}
\end{figure*}

The orientation of minijets is not correlated with the event geometry. They will also enhance energy deposition and entropy production. Consequently, hydrodynamic parameters need to be rescaled to match the data. In this study, we modify the overall normalization ($s_{\text{factor}}$) of the energy density after the IP-Glasma evolution and the constant shear viscosity to entropy density ratio ($\eta/s$) to asses the effects of minijets on medium evolution. The values of these parameters depend on the minimum possible transverse momentum of each parton in a back-to-back parton pair produced in a hard scattering ($p^{\text{J}}_{\text{min}}$), which is chosen to be above saturation scale. The optimum values of these parameters for three different choices of $p^{\text{J}}_{\text{min}}$ are fixed to reproduce charged hadron multiplicity and $v_2$ (see Fig.~\ref{fig:spectra_v2}). The obtained parameters are compared to the case without minijets in Table~\ref{tab:newparams}.

\begin{table}
    \centering
    \begin{tabular}{c|c|c}
      $p^{\text{J}}_{\text{min}}$  & $s_{\text{factor}}$ & $\eta/ s$  \\\hline
        4 GeV & 0.45 & 0.02 \\
        7 GeV & 0.82 & 0.1 \\
        10 GeV & 0.9 & 0.125 \\ 
        No Jets & 0.915 & 0.13 \\
    \end{tabular}
    \caption{Optimum values of initial state normalization and the shear viscosity to entropy density ratio for different $p^{\text{J}}_{\text{min}}$.}
    \label{tab:newparams}
\end{table}

\begin{figure*}
    \centering
    \includegraphics[width=1\textwidth]{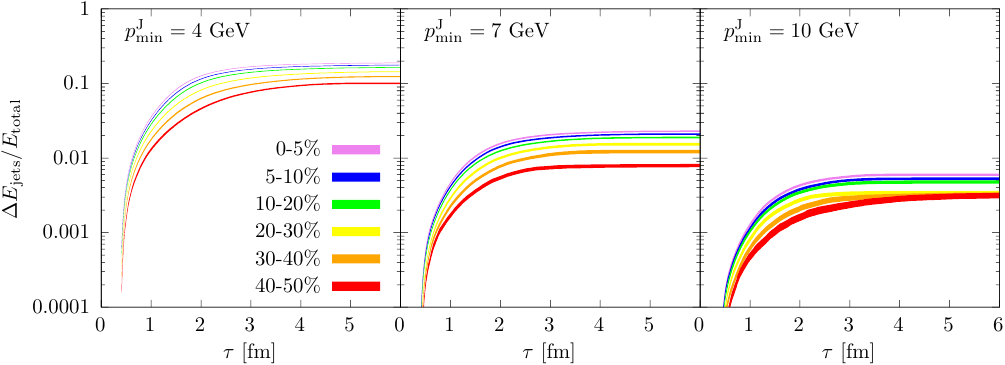}
    \caption{Ratio of the energy injected by minijets in the medium to the total energy of the medium as a function of proper time.}
    \label{fig:injected}
\end{figure*}

For $p^{\text{J}}_{\text{min}} = 4$ GeV, the value of $\eta/s$ and $s_{\text{factor}}$ need to be adjusted by about 85\% and 50\%, respectively.
This is because a sizeable portion of the energy in the bulk medium is being contributed by minijet sources, as seen in Fig.~\ref{fig:injected}. On top of this, numerous un-thermalized minijets, whose $x_{\rm stop}$ is longer than their path-length in QGP, also contribute to multiplicity. This ratio can be seen in Table~\ref{tab:fragnum}. The adjustments to these parameters for higher $p^{\text{J}}_{\text{min}}$ is less dramatic as the energy contribution from minijets decreases.

\begin{figure*}
    \centering
    \includegraphics[width=1.\textwidth]{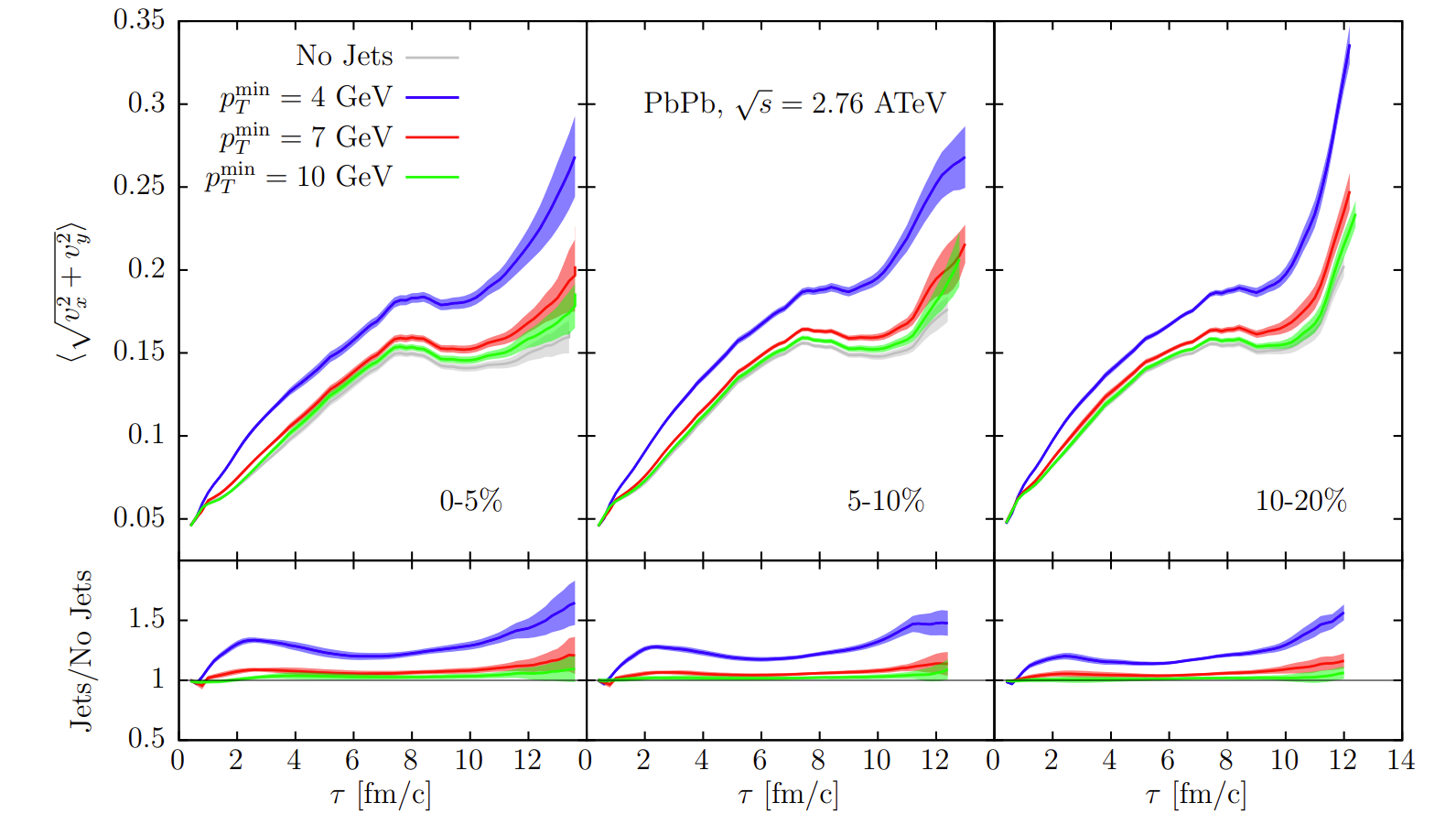}
    \caption{Averaged transverse fluid velocity as a function of time for different centralities.}
    \label{fig:flow}
\end{figure*}

\begin{table}
    \centering
    \begin{tabular}{c|c|c}
      $p^{\text{J}}_{\text{min}}$  & $\langle N_{\text{frag.}}/N_{\text{total}} \rangle_{0-5\%}$ & $\langle N_{\text{frag.}}/N_{\text{total}} \rangle_{40-50\%}$ \\\hline
        4 GeV & 0.077(1) & 0.252(3) \\
        7 GeV & 0.0125(5) & 0.033(2) \\
        10 GeV & 0.0042(3) & 0.014(2) \\
    \end{tabular}
    \caption{The average ratio of the number of hadrons coming from the fragmentation of un-thermalized partons to the total number of hadrons.}
    \label{tab:fragnum}
\end{table}

The flow profile is significantly different, even after accounting for changes in normalization and shear viscosity. Figure~\ref{fig:flow} shows the average transverse velocity for different choices of $p^\text{J}_{\text{min}}$. Flow develops much faster as more and more minijets deposit their momentum and enhance pressure gradients.

\begin{figure*}[t!]
    \centering
    \includegraphics[width=0.7\textwidth]{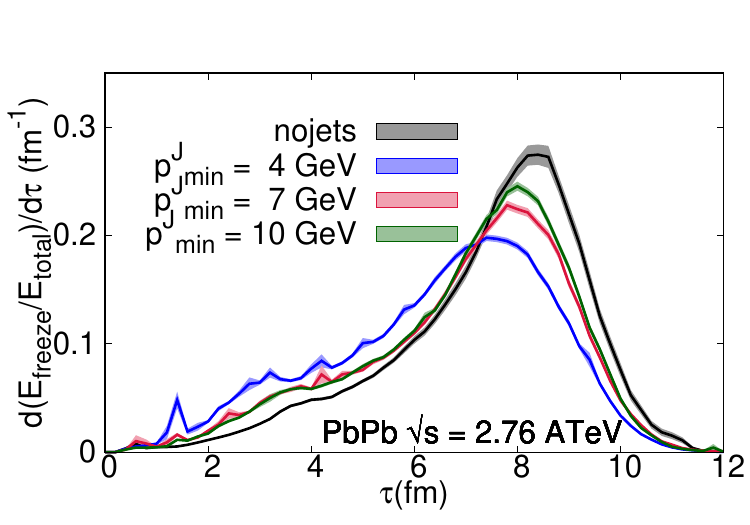}
    \caption{Fraction of energy being frozen out of a constant temperature hypersurface in 30-40\% centrality collisions, as a function of proper time.}
    \label{fig:frozen}
\end{figure*}

Minijets also modify the cooling of QGP fireball. They carry energy in their wake leaving cooler temperatures behind which break the constant temperature isotherms. This results in a larger portion of fireball freezing out earlier as evident in Figure~\ref{fig:frozen}.


\section{Discussion}\label{sec:results}
The simulations with minijets, with suitable adjustments to $s_{\text{factor}}$ and $\eta/s$,  reproduce the data well. The $p_T$-integrated spectra and $v_2$ are shown in Figure~\ref{fig:spectra_v2}. The differential spectra and $v_n$ can also be reasonable reproduced for different choices of $p^{\text{J}}_{\text{min}}$~\cite{Pablos:2022piv}.

The enhanced entropy from the minijets need a downward adjustment of the overall normalization factor of the initial energy density in the hydrodynamic medium. This is to ensure that the correct multiplicities are reproduced. The enhanced fluctuations also require re-tuning of the shear viscosity to entropy density ratio. This has important implications for extraction of QGP properties from model-to-data comparisons in the heavy-ion collision program. 

The size of these effects depend on the soft-hard separation scale $Q_s$. This was an open parameter in our study. In principle, this could be optimized from a systematic Bayesian study of this model and experimental data. However, it is more likely that the soft and hard modes do not cleanly separate at a particular scale, and there is an overlap between the two. This warrants an energy loss model with a gradual separation between the soft and hard modes. These aspects are left for future studies.

\section*{Acknowledgements}
This work was funded in part by the Natural Sciences and Engineering Research Council of Canada (C. G., S. J.) and in part by the U.S. DOE under Grant No. DE-FG02-87ER40328 (M. S.) Computations were made on the Beluga supercomputer
at McGill University, managed by Calcul Québec and by the Digital Research Alliance of Canada. D.P. has received funding from the European Union’s Horizon 2020 research and innovation program under the Marie Sklodowska-Curie grant agreement No. 754496.

\end{document}